\documentclass[journal,comsoc]{IEEEtran}
\usepackage{cite}
\usepackage{amsmath,amssymb,amsfonts}
\usepackage{algorithmic}
\usepackage{graphicx}
\usepackage{hyperref}

\usepackage{multirow}
\usepackage{booktabs,siunitx}
\usepackage{makecell}

\usepackage{balance}
\usepackage{flushend}

\usepackage{float}

\usepackage{graphicx}
\usepackage{longtable}

\begin{document}
\title{Collaborative Filtering-Based Method for Low Resolution and Details Preserving Image Denoising}
\author{
	\textsuperscript{1}Basit~O. Alawode, \textsuperscript{2}Mudassir Masood, \textsuperscript{3}Tarig Ballal, and \textsuperscript{4}Tareq Al-Naffouri.\\
	\textsuperscript{1,2}King Fahd University of Petroleum and Minerals, Dhahran, Saudi Arabia. \\
	\textsuperscript{3,4}King Abdullah University of Science and Technology, Thuwal, Saudi Arabia. \\
	Email: {\{\textsuperscript{1}g201707310, \textsuperscript{2}mudassir\}@kfupm.edu.sa, 
	\{\textsuperscript{3}tarig.ahmed, \textsuperscript{4}tariq.alnaffouri\}@kaust.edu.sa}
}
\maketitle

\begin{abstract} 
Over the years, progressive improvements in denoising performance have been achieved by several image denoising algorithms that have been proposed. Despite this, many of these state-of-the-art algorithms tend to smooth out the denoised image resulting in the loss of some image details after denoising. Many also distort images of lower resolution resulting in partial or complete structural loss. In this paper, we address these shortcomings by proposing a collaborative filtering-based (CoFiB) denoising algorithm. Our proposed algorithm performs weighted sparse-domain collaborative denoising by taking advantage of the fact that similar patches tend to have similar sparse representations in the sparse domain. This gives our algorithm the intelligence to strike a balance between image detail preservation and noise removal. Our extensive experiments showed that our proposed CoFiB algorithm does not only preserve the image details but also perform excellently for images of any given resolution where many denoising algorithms tend to struggle, specifically at low resolutions.
\end{abstract}


\section{INTRODUCTION}

For its practical applications in fields such as astronomy, medical, autonomous driving, etc., image denoising as an area of research has been around for a very long time. In spite of the advancements in technology which has made it possible to obtain crystal clear images in recent times, image capturing equipment have very little control over the environment in which such images are taken. For instance, medical images of human bones captured by X-ray machines have the various tissues and organs of the body as potential noise. This implies that noise in images cannot be totally removed by such equipment making image denoising a very active research area till today.

Image denoising is, therefore, the art of recovering a clean image $ \mathbf{X} $ from a noisy image observation $ \mathbf{Y} $. $ \mathbf{Y} $ is a combination of the clean image $ \mathbf{X} $ and a noise $ \mathbf{W} $ typically modeled as in \eqref{eq:noisyImage}. 

\begin{equation}
	\mathbf{Y = X + W.}
	\label{eq:noisyImage}
\end{equation}

The entries of $ \mathbf{W} $ are assumed to be random and independently drawn from the same probability distribution (i.i.d) such as the Gaussian distribution. As such, $ \mathbf{W} $ is typically modeled as an additive, white and Gaussian noise (AWGN) with zero mean and a known standard deviation $\sigma$.

Over the years, several image denoising algorithms have been proposed. Some of them exploit various mathematical analysis and similarities in pixels and patches to perform denoising. Such algorithms are commonly referred to as classical denoising algorithms, such as \cite{buades2005, Behzad2017, Aharon2006a, Dabov2007a, mairal2009}. The non-local means (NLM) algorithm \cite{buades2005} uses the similarities in the pixels of the corrupted image to perform denoising. Its performance, however, has been surpassed by the patch-based algorithms. This is because a patch carries more information about an image as compared to a single pixel. 

To perform patch-based denoising, the overlapping patches are usually transformed from the spatial domain to a transform domain. This transform domain is such that the noisy components are easily separable from the original image's components. Denoising then becomes a simple question of identifying which components are due to the noise and removing them. One of such domain which has been exploited by most patch-based denoising algorithms is the sparse domain. In this domain, the patches which are dense in the spatial domain are represented by a few nonzero components. This makes denoising easier and more effective. For such transformation, an overcomplete dictionary with the help of sparse recovery algorithms such as the orthogonal matching pursuit (OMP) \cite{Aharon2006a} and the support agnostic bayesian matching pursuit (SABMP) \cite{Masood2013a} algorithms are used. The overcomplete dictionary is obtained using analytical techniques such as the discrete cosine transform (DCT) and discrete wavelet transform (DWT) \cite{qayyum2016}, wavelet \cite{daubechies1990}, curvelet \cite{starck2002}, and contourlet \cite{do2005} dictionaries. Another approach to obtaining the dictionary is to learn it from the image patches themselves. This is done with the help of dictionary learning algorithms such as the K-SVD \cite{Aharon2006a}, the method of optimal directions (MOD) \cite{engan1999}, principal component analysis (PCA) \cite{vidal2003}, etc. Algorithms such as the Block-Matching and 3D transformation (BM3D) \cite{Dabov2007a}, Collaborative Support-Agnostic Recovery (CSAR) \cite{Behzad2017}, and K-SVD denoising \cite{Elad2006a} algorithms are some of a long list of algorithms which utilizes patches transformation into the sparse domain to perform denoising.  

Owing to its effectiveness, the BM3D was used as the benchmark denoising algorithm until recently. The use of neural networks (NNs) in denoising has seen the performance of the state-of-the-art classical techniques surpassed. The NN algorithms mostly utilize several layers of convolutional NNs (CNNs) to perform denosing. Example of such algorithms are the deep CNN (DnCNN) \cite{Zhang2017} and the fast and flexible feed-forward NN (FFDNet) \cite{Zhang2018}. 

Despite the progressive improvement in the performance of the denoising algorithms over the years, many of them fail in their inability to denoise images of lower resolutions. Noisy low resolution images are sometimes the output of medical imaging systems. As such, ability of the image denoising algorithms to denoise such images is of practical importance. Some of the algorithms also tend to smooth out the denoised image leading to the loss of some of the salient details of the image. 

In this paper, we present a collaborative filtering-based (CoFiB) denoising algorithm which performs collaboration among similar patches in the sparse domain to perform denoising. Our contribution is the ability of our proposed algorithm to perform well across images of any given resolution. Most importantly, images of low resolution. While denoising, our algorithm is also smart enough to preserve the details of the image. This means that our algorithm is able to strike a balance between complete image denoising leading to smoothened effect on the image and details preservation.

The rest of the paper is organized as follows. We provide a detail description of our proposed CoFiB algorithm in Section \ref{sec:cofib}. Section \ref{sec:exp} provides the details of the expensive experiments conducted and the results. Finally, we conclude in Section \ref{sec:con}.
 
\section{THE PROPOSED COLLABORATIVE FILTERING-BASED DENOISING ALGORITHM}
\label{sec:cofib}

\begin{figure*}
	\centering
	\includegraphics[height=9cm, width=\textwidth]{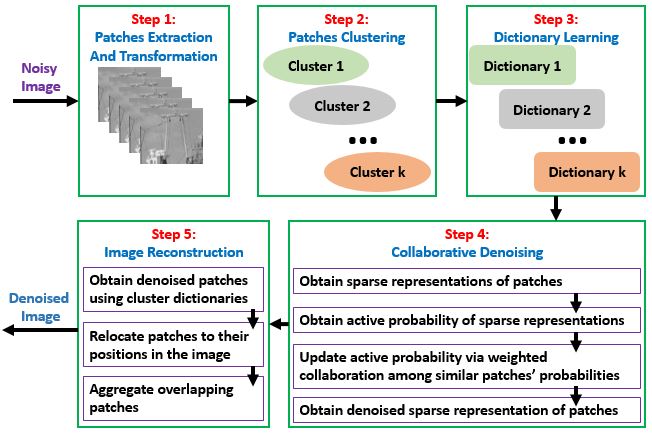}
	\caption{Flowchart of the Proposed CoFiB Algorithm.}
	\label{fig:cofibFlowchart}
\end{figure*}

In this section, we present the proposed CoFiB denoising algorithm. Our algorithm is a patch-based classical image denoising algorithm. The patches are transformed to the sparse domain where collaborative denoising is performed. In the sparse domain, the algorithm takes advantage of the similar patches' sparse representations to obtain a better denoising performance. The denoised sparse  representations are transformed back to the spatial domain. All denoised patches are then brought together to generate the denoised image. All these are performed in 5 steps as shown in Fig. \ref{fig:cofibFlowchart}. These steps are described as follows.

\subsection{Step 1: Patches Extraction and Transformation}
\label{sec:patchExtract}

The first step in the pipeline of the proposed CoFiB algorithm is to extract a set of overlapping patches from the noisy image. For ease of analysis, the patches are of squared sizes, i.e. having the same number of row and column pixels. The patches are formed in such a way that the number of patches is equal to the number of pixels in the noisy image. This is done by padding the image with $ \frac{\sqrt{R}-1}{2} $ pixels in all directions. Given that each patch is of size $ n \times n $, $ R = n^{2} $ is the size of each vectorized image patch. $ R $ is chosen in a way that $ n $ is always odd. The reason for choosing $ R $ in this manner is to ensure that $ \frac{\sqrt{R}-1}{2} $ will always produce an integer. This implies that all pixels in the image would have a corresponding patch, known as the pixel patch. The central pixel of a pixel patch corresponds to the pixel it is representing. 

\begin{figure}[htb]
	\centering
	\begin{center}
		\begin{tabular}{ c c c }
			\includegraphics[width= 3cm, height= 3cm]{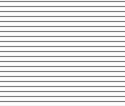}
			& 
			$ \qquad $
			&
			\includegraphics[width= 3cm, height= 3cm]{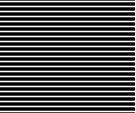}
			\\
			(A) & & (B)
		\end{tabular}
	\end{center}
	\caption{Patches with Similar Pattern}
	\label{fig:simpatchesexample}
\end{figure}

Some parts of both natural and man-made images have similar patterns that may or may not differ in intensity levels. For example, Fig. \ref{fig:simpatchesexample}(A) and Fig. \ref{fig:simpatchesexample}(B) have different intensity levels, but the underlying pattern/structure is similar. Similarly, a purely white image and a purely black image can be thought of as having the same pattern but, while the intensity value of a white image is 0, that of a black is 255 (or 0 for white and 1 for black for an image range of [0, 1]). In our algorithm, we want to ensure that all patches that have a similar structure are grouped together in an intensity invariant manner. This was achieved by dividing a patch by the maximum absolute pixel value. This ensure that patches are grouped together not by their intensity but by their underlying structural pattern. 

\subsection{Step 2: Patches Clustering}
\label{sec:patchCluster}

Our proposed algorithm performs collaboration in the sparse domain. To transform the intensity invariant patches from the spatial to the sparse domain, an overcomplete dictionary is needed. However, because the patches are coming from different parts of the image, they will be very different. Therefore, they will be scattered in the sparse domain. For collaboration, we need patches close enough to each other in the sparse domain. Similar patches are expected to have similar representations in the sparse domain. We take advantage of this fact and grouped similar intensity invariant patches together. These groupings are used to enhance the sparse domain collaborative denoising. The popular k-means clustering algorithm \cite{jain2010}  was used to group the intensity invariant patches ointo $ k $ clusters.

\subsection{Step 3: Dictionary Learning}
\label{sec:dictLearn}

The dictionary for each cluster was learned with the help of the K-SVD dictionary learning algorithm \cite{Aharon2006}. For each cluster, we learned a separate dictionary. While a little computationally expensive, this helped to ensure that each cluster is suitably represented and most of the sparse estimation components better project the patches of each cluster unto their corresponding sparse domain with negligible error compared to using a single dictionary for all the patches.

\subsection{Step 4: Collaborative Denoising}
\label{sec:collabDen}

The sparse representation of each patch in each cluster is then obtained using the Support Agnostic Bayesian Matching Pursuit (SABMP) sparse recovery algorithm \cite{Masood2013a}. Although there are several sparse recovery algorithms, our choice of the SABMP algorithm hinged on the following facts about the algorithm which helped in the strategy we employed for denoising the image.

\begin{enumerate}
	\item It is invariant to the distribution of the image intensity values.
	
	\item It does not impose any strict conditions on the cluster dictionary,
	
	\item and it gives, as an output, the probability of each element of the sparse representation of a patch to be active.
\end{enumerate}

The sparse representation of patches in a cluster should have a high number of common locations of nonzero elements provided that the dictionary used is perfect, that is, all the dictionary atoms are uncorrelated. This would mean that the uncommon nonzero elements in the sparse representation are due mainly to noise, to a high probability. By simply removing the uncommon nonzero location components will leave us with clean portions of the representation. However, because the dictionary of a cluster was trained from similar data, a perfect dictionary is not obtainable. In addition, the dictionary is overcomplete, as such, several sparse domain representations could be mapped to a patch. This gives rise to sparse domain representations that are not unique. This ultimately means that several atoms in the dictionary can combine in different ways to yield the same point in the sparse domain. Locations of nonzero elements could differ due to genuine differences in the patches themselves. This makes simple removal of uncommon elements, not a good move. The SABMP algorithm comes to our rescue in this case as it can take into consideration the apriori probability of an element in a patch to be active in order to estimate the sparse vector. To take advantage of this fact, patches in the same cluster are allowed to collaborate to update their active probabilities in a weighted average approach to yield the denoised sparse representation's nonzero locations' probability. This step is as given mathematically in equations (\ref{eq:alpha}) through (\ref{eq:lambdaupdate}) below. 

\begin{subequations}
	\begin{align}
	\alpha_{j,p,k}=\dfrac{1}{\Delta(y_{j,k}, y_{p,k})} \quad j \neq p.
	\label{eq:alpha}
	\\[20pt]
	Normalized \qquad \alpha_{j,p,k}=0.5 \times \frac{\alpha_{j,p,k}}{\sum_{j=1}^{t-1}\alpha_{j,p,k}} \quad j \neq p.
	\label{eq:alphanormalized}
	\\[20pt]
	\lambda_{p,k}^{updated}=\sum_{j=1}^{t}{\alpha_{j,p,k}\lambda_{j,k}}, \quad \alpha_{t,p,k} = 0.5, \quad \lambda_{t,k} = \lambda_{p,k}.
	\label{eq:lambdaupdate} 
	\end{align}
\end{subequations}

For each patch $ p $ in each cluster $ k $, $ t-1 $ similar patches are selected (where $ t $ is the number of patches for collaboration). The selection is based on how close the patches in the cluster $ k $ are to the patch $ p $. This can be obtained using any suitable distance measure. A common distance metric is the Euclidean distance between the patch $ p $ (given as $ y_{p,k} $) and other patches in the same cluster $ k $ (given as $ y_{j,k} $, for $ j = 1 $ to $ n_{cluster} $, where $ n_{cluster} $ is the number of patches in the cluster). This distance is given as $ \Delta(y_{j,k}, y_{p,k}) $. The $ t-1 $ similar patches obtained are therefore given as $ y_{j,k} $, for $ j = 1 $ to $ t-1 $.  

The closer a patch is to the patch $ p $, the smaller the value of the distance between them, but the more similar they become. As such, a similar patch's denoising contribution to the patch $ p $ is inversely proportional to the distance between them. Hence, the denoising collaborative contribution ($ \alpha_{j,p,k} $) of a similar patch to the patch $ p $ in the same cluster $ k $ is obtained using equation(\ref{eq:alpha}). Observe the inverse proportionality in the equation. The distance of patch $ p $ to itself is $ 0 $, as such it should have the highest collaborative contribution to itself. This implies that patch $ p $ would have a 50\% normalized contribution value to itself, i.e. $ \alpha_{t,p,k} = 0.5 $. All the other similar patches would have to scale accordingly to contribute the other normalized 50\% value. This 50\%  normalized collaborative contribution value of the other similar patch to patch $ p $ is given by equation (\ref{eq:alphanormalized}). Observe how their contribution scale using $ 0.5 \times $ inverse of their summation as a scaling factor. This is the contribution of each obtained similar patch towards denoising the patch $ p $.

As earlier described, from the outputs of the SABMP algorithm, the location probabilities $ \lambda_{j,k} $ of the elements of the sparse representation of each of these similar patches including the patch $ p $ can be recovered. To obtain the denoised active probabilities of patch $ p $, its $ \lambda_{p,k} $ needs to be updated/denoised. The updated $ \lambda_{p,k}^{updated} $ is obtained using the normalized denoising collaborative contributions ($ \alpha_{j,p,k} $) of each of the patches with their respective $ \lambda_{j,k} $. This is as given in equation (\ref{eq:lambdaupdate}).

\begin{figure*}
	\centering
	\begin{center}
		\begin{tabular}{ c }
			\includegraphics[width=0.8\textwidth, height=0.4\textheight]{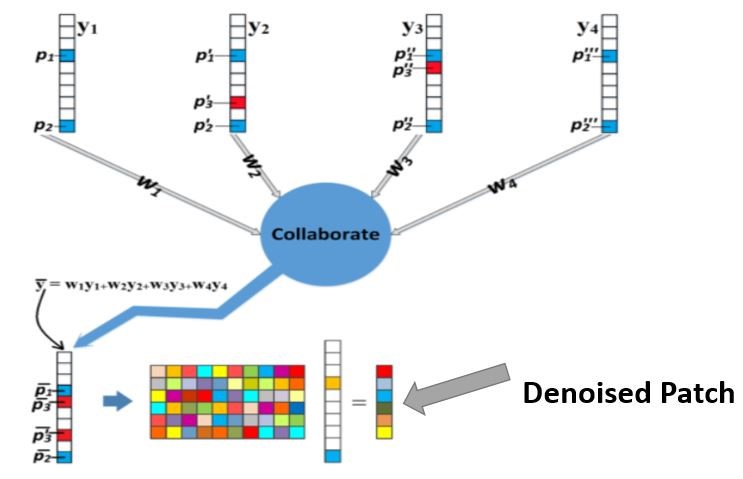}
			\\
			It should be noted here that y represents the active location \\
			probabilities of the sparse representation, $\lambda$, of a patch \\
			and w is the same as the collaboration weight $\alpha$ in (\ref{eq:alphanormalized}). 
		\end{tabular}
	\end{center}
	\caption{Collaborative Denoising of Patches in the Sparse Domain.}
	\label{fig:cofibcollab}
\end{figure*}

In Fig. \ref{fig:cofibcollab}, we show how the similar patches collaborate to update/denoise the location probability of the sparse representation of the currently considered noisy patch in each cluster. This updated/denoised probability is then fed back into the SABMP algorithm to re-estimate the sparse domain representation of the denoised patch. 

\subsection{Step 5: Image Reconstruction}
\label{sec:imageRecon}

The denoised sparse representation of the patches are then transformed back to the spatial domain. This is obtained by simply multiplying the denoised sparse representation of each patch with the dictionary of the cluster to which the patch belongs. This yields the denoised patches.

In order to reconstruct the denoised image, the denoised patches are simply replaced back to their original locations in the noisy image. Noting that these patches overlap, they are then aggregated to obtain the denoised image.

\section{EXPERIMENTAL RESULTS}
\label{sec:exp}

We performed several experiments on the proposed CoFiB algorithm. We also compared it to some of the classical denoising algorithms. For performance comparison, the widely used peak signal to noise ratio (PSNR) and the structural similarity index (SSIM) performance metrics have been used. These methods are termed objective measures. Despite these measures being widely used, visual inspection (subjective measure) of the denoised image is needed to truly ascertain the performance of an image denoising algorithm \cite{Al-Najjar2012}. We will show that in almost all cases, our proposed CoFiB algorithm performs subjectively (visually) better than most of the state-of-the-art denoising algorithms such as the K-SVD \cite{Elad2006a}, NLM \cite{buades2005}, and BM3D \cite{Dabov2007a}. However, using the objective metrics, some of the algorithms performed better in few cases. As such, an image denoising algorithm should be evaluated both subjectively and objectively to truly ascertain the performance of such algorithm.

To cluster the intensity invariant patches, we found $ k = 5 $ to be a good number to have yielded the best performance for our algorithm. This number can be explained by the fact that a natural image patch mostly consists of flat, horizontal, vertical, left diagonal, and right diagonal regions. We now proceed to present and describe the result of our experiments in the following sections.

\subsection{Performance Comparison of Different Algorithms on Different SNR}
\label{sec:cofibresultsdiffsnr}

\begin{figure*}
	\centering
	\includegraphics[width=\textwidth]{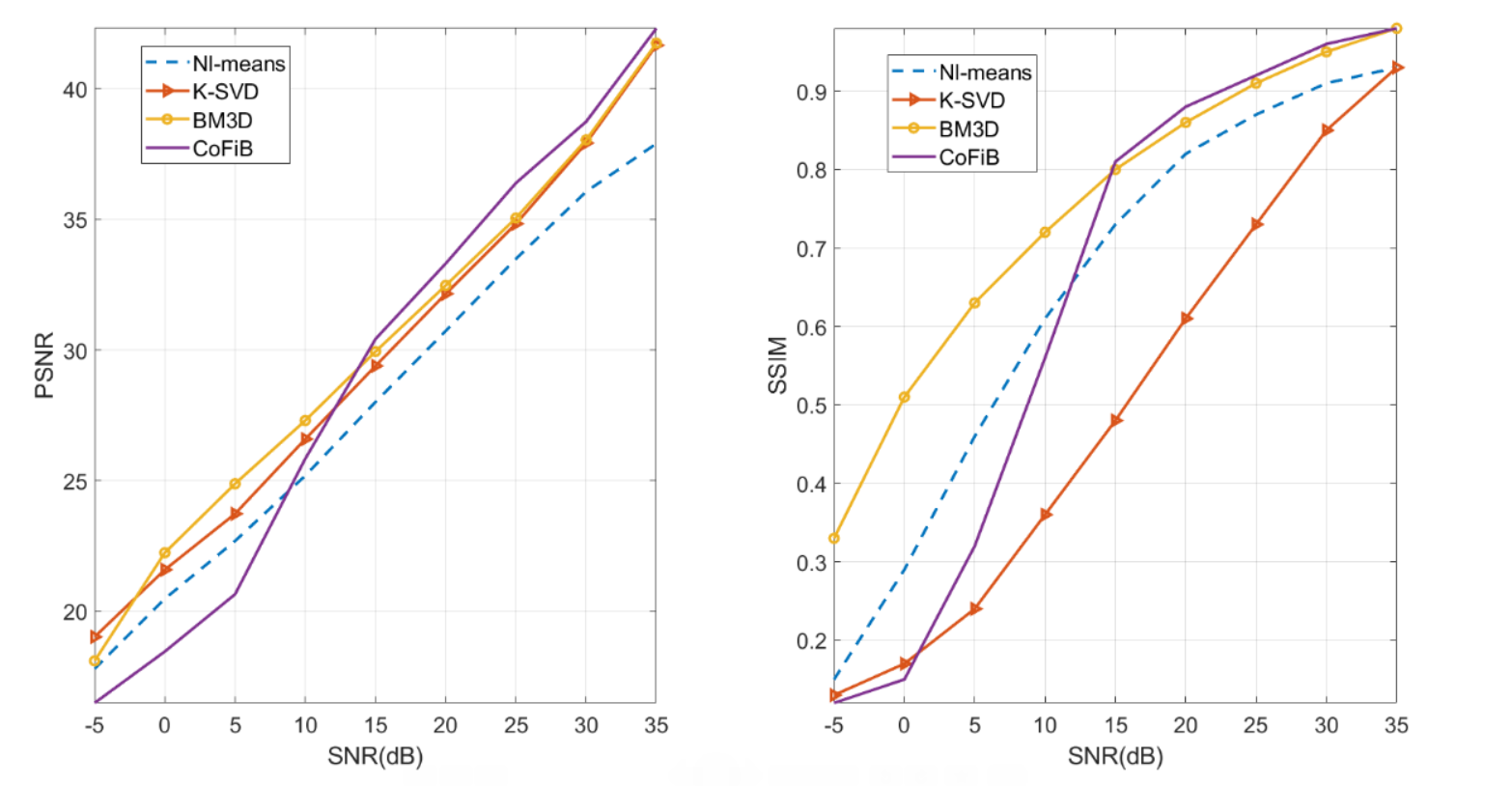}
	\caption{Performance Comparison of Different Algorithms w.r.t the SNR (Boat $ 512 \times 512 $)}
	\label{fig:cofibdiffsnr}
\end{figure*}

Here, we used a $ 512\times512 $ boat image which is one of the widely used images in literature for testing image denoising algorithms. White Gaussian Noise (WGN) was added to the original boat image. The noise was assumed to be of zero mean and variance that corresponds to the different values of SNR that were selected for our experiments. The SNR have been given in decibel (dB). These values ranges from $ -5dB $ to $ 35dB $ with an interval of $ 5dB $. We use the K-SVD, NLM, and BM3D to compare our algorithm. Each algorithm was then used to denoise the noisy image. Table \ref{tab:cofibdiffsnr} shows a comparison of the performance of the different algorithms on the different SNR values. These results are also presented graphically as shown in Fig. \ref{fig:cofibdiffsnr}.

\begin{table*}
	\centering
	\caption{Performance Comparison of Different Algorithms w.r.t the SNR (Boat $ 512 \times 512 $)}
	\begin{tabular}{ c }
		\includegraphics[width= \textwidth]{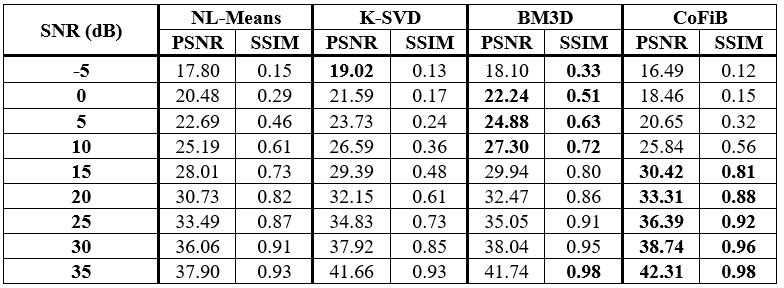}	
	\end{tabular}
	\label{tab:cofibdiffsnr}
\end{table*}

\begin{figure*}[htb]
	\centering
	\fbox{
		\begin{minipage}{13 cm}
			\begin{center}
				\begin{tabular}{ c c }
					\includegraphics[width= 0.4\textwidth]{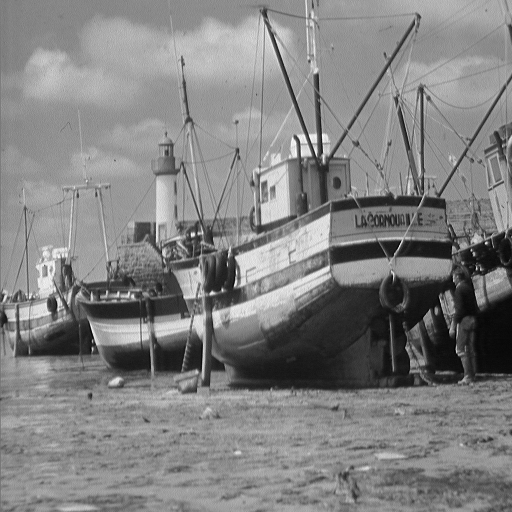} & \includegraphics[width= 0.4\textwidth]{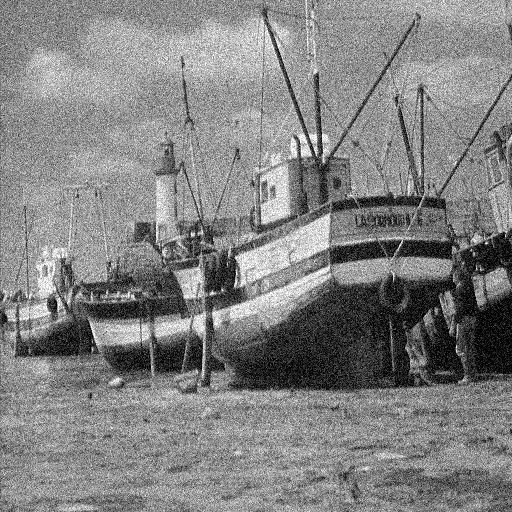}
					\\
					(a) & (b)
					\\
					\\
					\includegraphics[width= 0.4\textwidth]{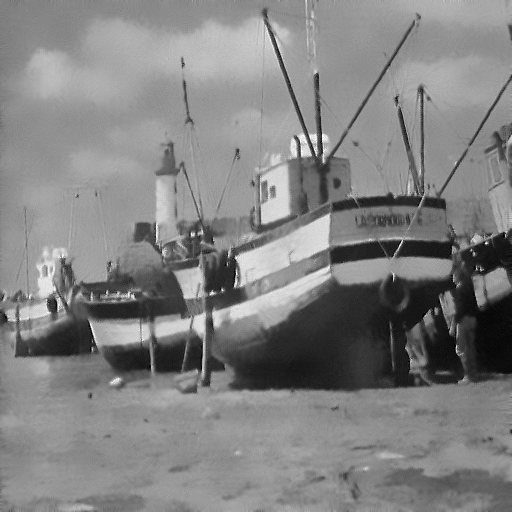} & \includegraphics[width= 0.4\textwidth]{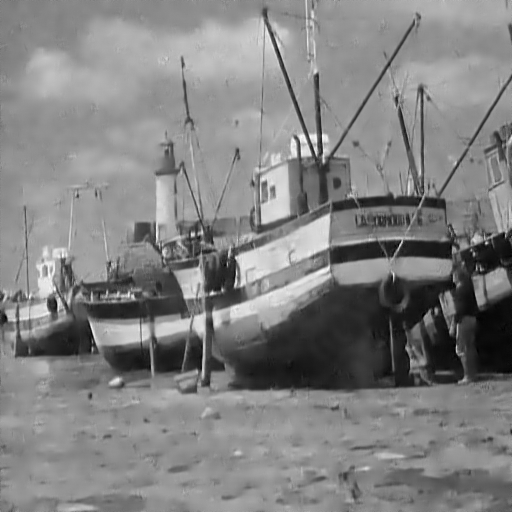}
					\\
					(c) & (d)
					\\
					\\
					\includegraphics[width= 0.4\textwidth]{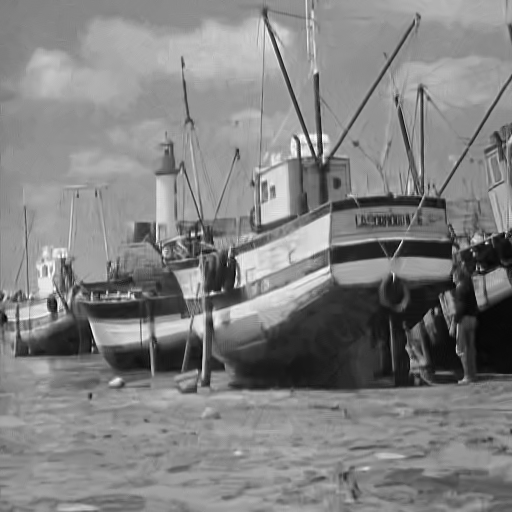} & \includegraphics[width= 0.4\textwidth]{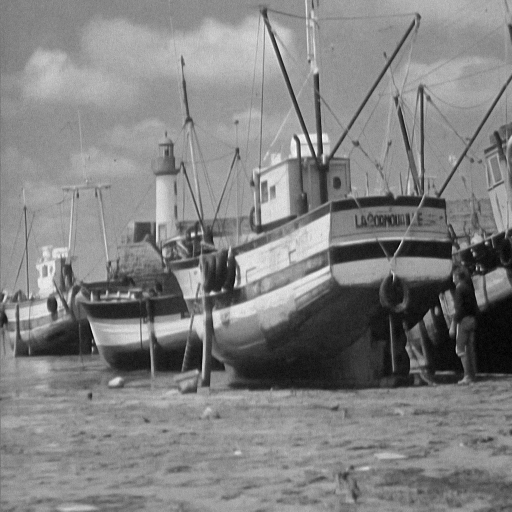}
					\\
					(e) & (f)
					\\
					\\
					\multicolumn{2}{c}{(a) Clean Image, (b) Noisy Image, (c) NLM} \\
					\multicolumn{2}{c}{(d) K-SVD, (e) BM3D, (f) CoFiB}
				\end{tabular}
			\end{center}
		\end{minipage}
	}
	\caption{Comparison of Results on Boat Image of SNR 20dB.}
	\label{fig:cofibdiffsnrimages}
\end{figure*}

It can be noticed from Table \ref{tab:cofibdiffsnr} and Fig. \ref{fig:cofibdiffsnr} that the CoFiB algorithm performed better than the results of some of the other state-of-the-art algorithms, except for severe noise level. Denoised images for SNR of 20dB are also presented in Fig. \ref{fig:cofibdiffsnrimages}. Emphasis should be placed on the flat regions and how edges are better preserved in CoFiB. It should also be observed in Fig. \ref{fig:cofibdiffsnrimages} that the visual performance of the CoFiB algorithm is closer to the original image compared to that of the other algorithms.

For the NLM result shown in Fig. \ref{fig:cofibdiffsnrimages}(c), the areas around the sand were smooth out, despite producing a PSNR of 30.73 and SSIM of 0.82. Also, K-SVD seems to produce a better result (Fig. \ref{fig:cofibdiffsnrimages}(d)) but its PSNR is lower than that of BM3D in Fig. \ref{fig:cofibdiffsnrimages}(e). The result of BM3D in Fig. \ref{fig:cofibdiffsnrimages}(e) was also smooth out compared to that of the proposed CoFiB algorithm. The CoFiB’s result, Fig. \ref{fig:cofibdiffsnrimages}(f), can clearly be observed to have preserved all the details in the image, despite objective performance sometimes claiming otherwise. This detail preserving nature of our proposed algorithm would become more apparent when we compare the different algorithms on images of varying resolution in the next section.

\subsection{Performance Comparison of Different Algorithms on Different Image Resolutions}
\label{cofibresultsdiffresolution}

In this section, we compare the performance of the denoising algorithms for various image resolutions. We, again, selected the boat image for this purpose. We generated images of three different resolutions of $ 64\times64 $, $ 128\times128 $, and $ 256\times256 $ from the image. WGN was added to these images with SNR of 20dB and then denoised using the different algorithms. Table \ref{tab:cofibdiffresolution} shows the results of the various algorithms on the different boat image resolutions. It can be clearly observed that the proposed CoFiB algorithm performs objectively (in terms of PSNR and SSIM) better than all the other algorithms. 

In addition, Table \ref{tab:cofibdiffresolutionresult} shows the denoised images of the various resolutions that were considered. From this table, it can clearly be seen that for lower resolution images of $ 64\times64 $ and $ 128\times128 $, the denoised output was completely distorted by all the other algorithms except our proposed CoFiB algorithm. CoFiB at the same time preserved the edges and flat regions of the image which the other algorithms tend to find a challenge at achieving. CoFiB gave clearer, sharper, and undistorted denoised images. This is a facr that we earlier highlighted that many algorithms find it challenging to denoise images of lower resolutions. Our proposed CoFiB algorithm can clearly be observed to have performed reasonably well for any given image resolution. 

An ideal image denoising algorithm should work well for any given image. For this reason, in the next section, we present the comparison of our proposed algorithm with the other algorithms for different images. 

\begin{table*}
	\centering
	\caption{Performance of Different Algorithms on Different Resolutions (Boat, SNR of 20dB)}
	\begin{tabular}{ c }
		\includegraphics[width= \textwidth]{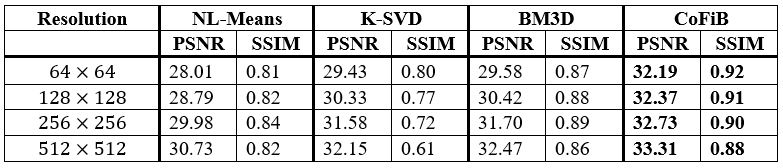}	
	\end{tabular}
	\label{tab:cofibdiffresolution}
\end{table*}

\begin{table*}
	\centering
	\caption{Performance Comparison of Different Algorithms on Different Images (SNR = 20dB)}
	\begin{tabular}{ c }
		\includegraphics[width= \textwidth]{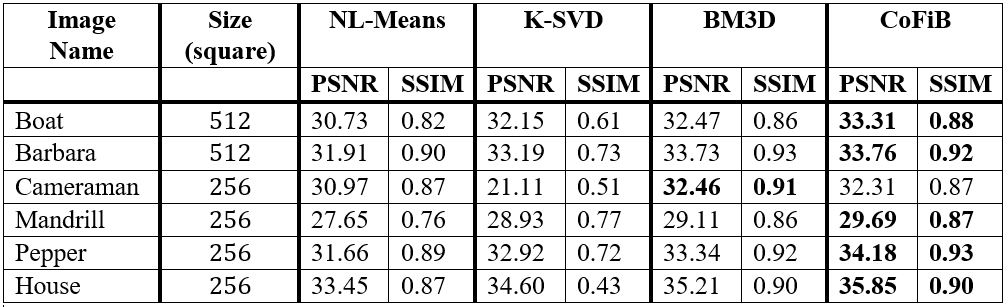}	
	\end{tabular}
	\label{tab:cofibdiffimages}
\end{table*}

\subsection{Performance Comparison of Different Algorithms on Different Images}
\label{cofibresultsdiffimages}

Here, we compare our proposed algorithm with the other algorithms on different images. The images used are some of the commonly used ones in literature to test denoising algorithms. In this paper, we have shown the results for 6 different images namely the boat, barbara, cameraman, mandrill, pepper and house images. For visual inspection of the performance, we have shown the results of the boat, cameraman, and the mandrill images. 

Table \ref{tab:cofibdiffimages} shows the objective comparison of the performance of the different algorithms on the considered images. Similar to other experiments, WGN of fixed SNR of 20dB was added to the original images. The various algorithms are then used to denoise the images. From Table \ref{tab:cofibdiffimages}, it can easily be observed that our proposed CoFiB algorithm performed better than all the other algorithms both in terms of PSNR and SSIM except for the Cameraman image where the BM3D algorithm performed slightly better. 

Furthermore, in Table \ref{tab:cofibdiffimagesresult}, we show three original and noisy images along with their denoised versions for all the selected algorithms for subjective performance comparison. It can be observed from this table that the proposed CoFiB result presented a more realistic denoised version of the original image. While results are close visually, differences can still be observed along the nose of the mandrill image.

\section{CONCLUSION}
\label{sec:con}

In this paper, a collaborative filtering-based method for image denoising was proposed where collaboration among the sparse representations of similar patches in the sparse domain was adopted to perform denoising. These sparse domain collaboration helped in improving the denoising capability of the algorithm. Unlike most of the other algorithms that tend to smooth out the denoised image resulting in loss of salient image details, our proposed CoFiB algorithm was able to preserve the details of the denoised image. This was achieved by the ability of the algorithm to strike a balance between noise removal and detail preservation. Additionally, most denoising algorithms tend to completely distort images at relatively lower resolutions. Our proposed algorithm was able to conveniently preserve the structure of the image for all the image resolutions considered. Extensive experimental results demonstrated that our proposed algorithm does not only performed objectively well (in terms of PSNR and SSIM) but also performed subjectively (i.e. visually) better. In future, we intend to translate our proposed algorithm into a deep learning denoising algorithm.  

\section{ACKNOWLEDGEMENT}
\label{sec:ack}

We gratefully acknowledge the support of the KAUST Supercomputing Lab for providing us with the computing cluster (ibex) that were used to carry out this research.
\balance

\clearpage
\onecolumn
\begin{center}
	\begin{longtable}{c c c}
		\caption{Results of Different Image Resolutions.} 
		\label{tab:cofibdiffresolutionresult} \\
		
		\hline
		\hline 
		\multicolumn{1}{c}{\textbf{$ 64 \times 64 $}} & \multicolumn{1}{c}{\textbf{$ 128 \times 128 $}} & \multicolumn{1}{c}{\textbf{$ 256 \times 256 $}} \\ \hline 
		\endfirsthead
		
		\multicolumn{3}{c}%
		{{\bfseries \tablename\ \thetable{} -- continued from previous page}} \\
		\hline \multicolumn{1}{c}{\textbf{$ 64 \times 64 $}} &
		\multicolumn{1}{c}{\textbf{$ 128 \times 128 $}} &
		\multicolumn{1}{c}{\textbf{$ 256 \times 256 $}} \\
		\hline \\
		\endhead
		
		\multicolumn{3}{r}{{Continued on next page}} \\
		\endfoot
		
		\hline \hline
		\endlastfoot
		
		Original & Original & Original\\
		\includegraphics[width= 0.3\textwidth]{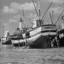} &	
		\includegraphics[width= 0.3\textwidth]{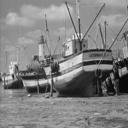}
		&
		\includegraphics[width= 0.3\textwidth]{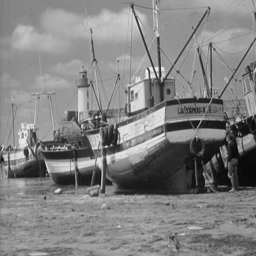}
		\\
		\hline
		Noisy & Noisy & Noisy \\
		\includegraphics[width= 0.3\textwidth]{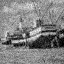} &	
		\includegraphics[width= 0.3\textwidth]{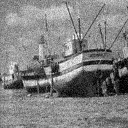}
		&
		\includegraphics[width= 0.3\textwidth]{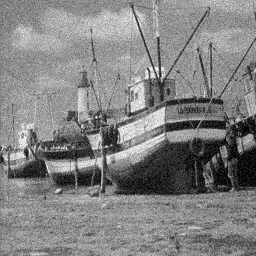}
		\\
		\hline
		NLM & NLM & NLM\\
		\includegraphics[width= 0.3\textwidth]{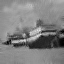} &	
		\includegraphics[width= 0.3\textwidth]{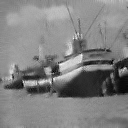}
		&
		\includegraphics[width= 0.3\textwidth]{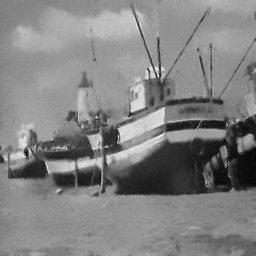}
		\\
		\hline
		K-SVD & K-SVD & K-SVD\\
		\includegraphics[width= 0.3\textwidth]{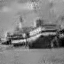} &	
		\includegraphics[width= 0.3\textwidth]{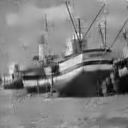}
		&
		\includegraphics[width= 0.3\textwidth]{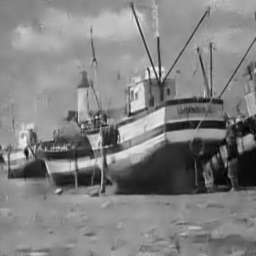}
		\\	
		\hline
		BM3D & BM3D & BM3D\\
		\includegraphics[width= 0.3\textwidth]{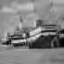} &	
		\includegraphics[width= 0.3\textwidth]{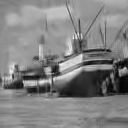}
		&
		\includegraphics[width= 0.3\textwidth]{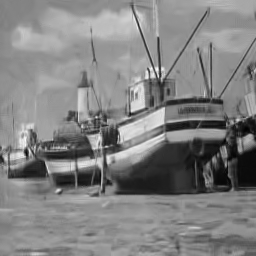}
		\\
		\hline
		CoFiB & CoFiB & CoFiB\\
		\includegraphics[width= 0.3\textwidth]{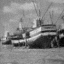} &	
		\includegraphics[width= 0.3\textwidth]{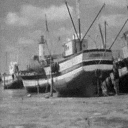}
		&
		\includegraphics[width= 0.3\textwidth]{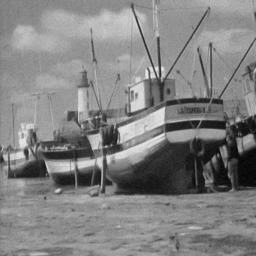}
		\\
	\end{longtable}
\end{center}

\newpage
\begin{center}
	\begin{longtable}{c c c}
		\caption{Results of Some Selected Images (SNR of 20dB)} 
		\label{tab:cofibdiffimagesresult} \\
		
		\hline
		\hline 
		\multicolumn{1}{c}{\textbf{Boat Image}} & \multicolumn{1}{c}{\textbf{Cameraman Image}} & \multicolumn{1}{c}{\textbf{Mandrill Image}} \\ 
		$ 512 \times 512 $ & $ 256 \times 256 $ & $ 256 \times 256 $ \\ \hline 
		\endfirsthead
		
		\multicolumn{3}{c}%
		{{\bfseries \tablename\ \thetable{} -- continued from previous page}} \\
		\hline \multicolumn{1}{c}{\textbf{Boat Image}} &
		\multicolumn{1}{c}{\textbf{Cameraman Image}} &
		\multicolumn{1}{c}{\textbf{Mandrill Image}} \\
		\hline
		\\
		\endhead
		
		\multicolumn{3}{r}{{Continued on next page}} \\
		\endfoot
		
		\hline \hline
		\endlastfoot
		
		Original & Original & Original \\
		\includegraphics[width= 0.3\textwidth, height=0.3\textwidth]{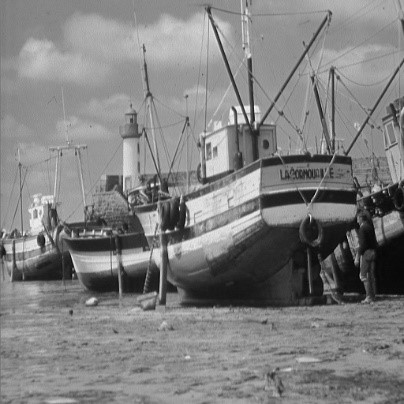} &	
		\includegraphics[width= 0.3\textwidth, height=0.3\textwidth]{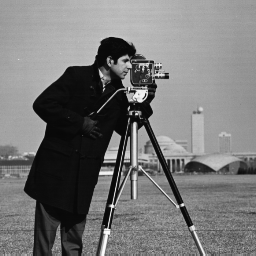}
		&
		\includegraphics[width= 0.3\textwidth, height=0.3\textwidth]{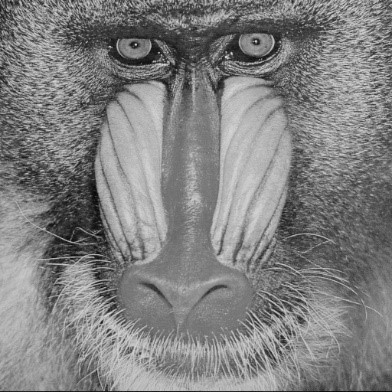}
		\\
		\hline
		Noisy & Noisy & Noisy \\
		\includegraphics[width= 0.3\textwidth, height=0.3\textwidth]{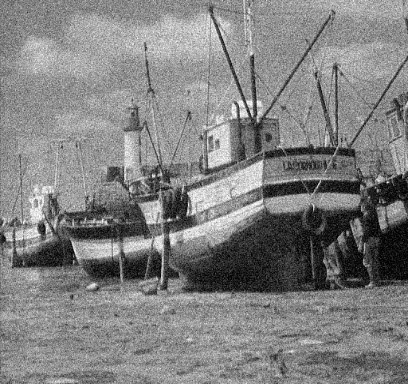} &	
		\includegraphics[width= 0.3\textwidth, height=0.3\textwidth]{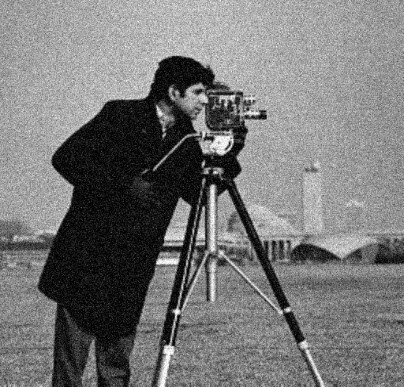}
		&
		\includegraphics[width= 0.3\textwidth, height=0.3\textwidth]{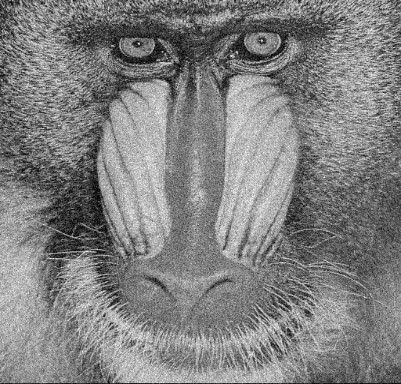}
		\\
		\hline
		NLM & NLM & NLM\\
		\includegraphics[width= 0.3\textwidth, height=0.3\textwidth]{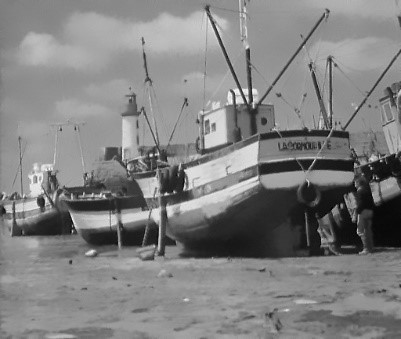} &	
		\includegraphics[width= 0.3\textwidth, height=0.3\textwidth]{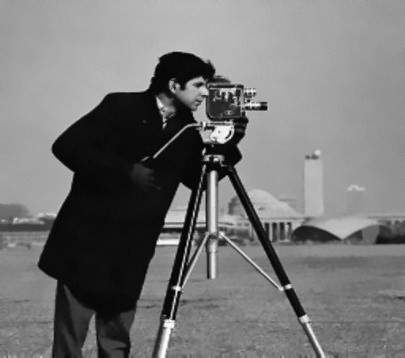}
		&
		\includegraphics[width= 0.3\textwidth, height=0.3\textwidth]{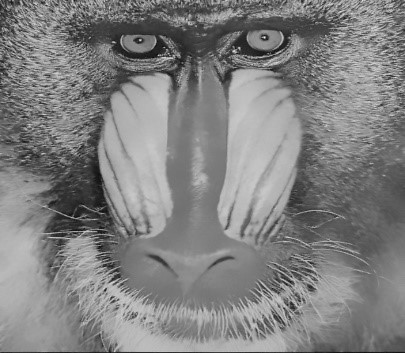}
		\\
		\hline
		K-SVD & K-SVD & K-SVD\\
		\includegraphics[width= 0.3\textwidth, height=0.3\textwidth]{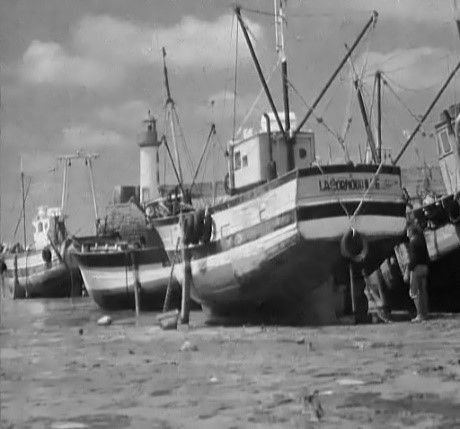} &	
		\includegraphics[width= 0.3\textwidth, height=0.3\textwidth]{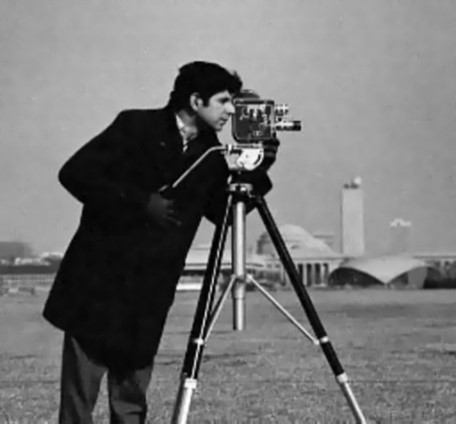}
		&
		\includegraphics[width= 0.3\textwidth, height=0.3\textwidth]{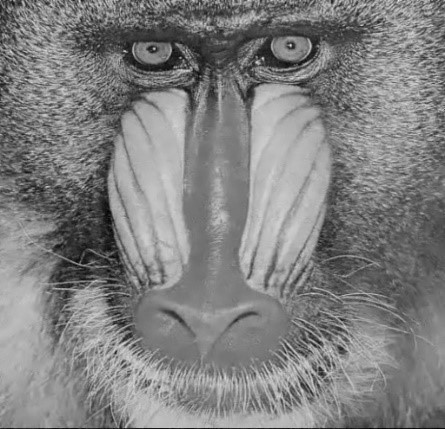}
		\\	
		\hline
		BM3D & BM3D & BM3D\\
		\includegraphics[width= 0.3\textwidth, height=0.3\textwidth]{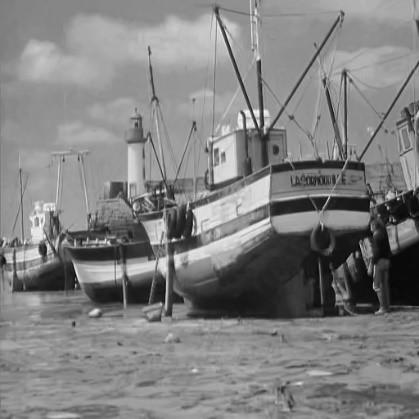} &	
		\includegraphics[width= 0.3\textwidth, height=0.3\textwidth]{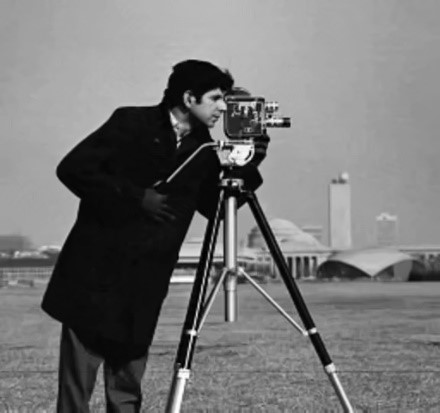}
		&
		\includegraphics[width= 0.3\textwidth, height=0.3\textwidth]{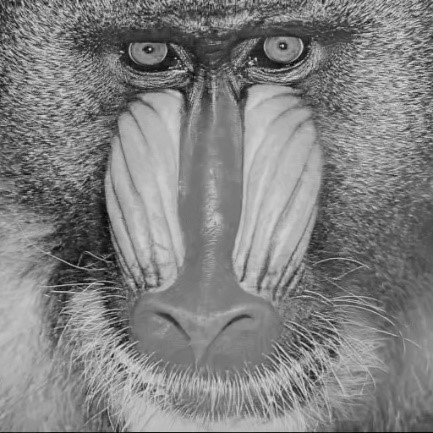}
		\\
		\hline
		CoFiB & CoFiB & CoFiB\\
		\includegraphics[width= 0.3\textwidth, height=0.3\textwidth]{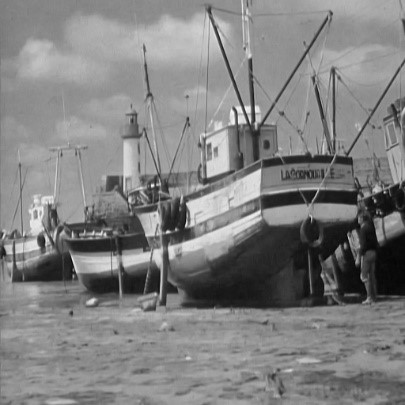} &	
		\includegraphics[width= 0.3\textwidth, height=0.3\textwidth]{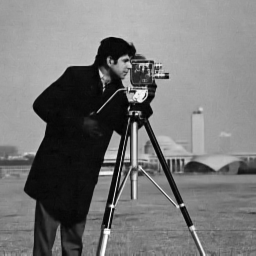}
		&
		\includegraphics[width= 0.3\textwidth, height=0.3\textwidth]{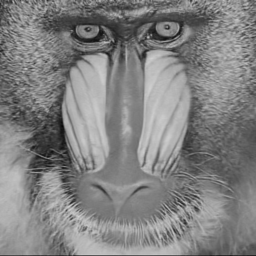}
		\\
	\end{longtable}
\end{center}
\clearpage
\twocolumn

\bibliographystyle{IEEEtran}
\bibliography{/Users/BASTECH/Documents/Mendeley/KAUST}

\end{document}